\documentclass[conference,twoside,a4paper]{IEEEtran}
\makeatletter
\def\@oddhead{\hbox{}\scriptsize\rightmark \hfil \thepage}
\def\@evenhead{\scriptsize\thepage \hfil \leftmark\hbox{}}
\makeatother
\footernote{M. Grassl and M. R\"otteler: QMDS Codes over Small Fields}

\usepackage{cite}

\ifCLASSINFOpdf
\else
\fi
\usepackage[cmex10]{amsmath}
\usepackage{amstext}
\usepackage{amssymb}

\IEEEoverridecommandlockouts

\newtheorem{theorem}{Theorem}
\newtheorem{conjecture}[theorem]{Conjecture}
\newtheorem{corollary}[theorem]{Corollary}
\newtheorem{definition}[theorem]{Definition}
\newtheorem{lemma}[theorem]{Lemma}
\newtheorem{proposition}[theorem]{Proposition}
\newtheorem{remark}[theorem]{Remark}

\usepackage{boldmath}
\usepackage{color}

\usepackage{epsf}
\usepackage{graphicx}

\def\QECC(#1,#2,#3,#4){[\![#1,#2,#3]\!]_{#4}}
\def\C{\mathbb{C}}
\def\F{\mathbb{F}}

\def\wgt{\mathop{\rm wgt}\nolimits}

\begin{document}
%
\title{Quantum MDS Codes over Small Fields\vskip-1ex}

\author{\IEEEauthorblockN{Markus Grassl}
\IEEEauthorblockA{
Universit\"at Erlangen-N\"urnberg\\
\& Max-Planck-Institut f\"ur die Physik des Lichts\\
Erlangen, Germany\\
Markus.Grassl@mpl.mpg.de}
\and
\IEEEauthorblockN{Martin R\"otteler}
\IEEEauthorblockA{Quantum Architectures and Computation Group\\
Microsoft Research\\
Redmond, WA, USA\\
martinro@microsoft.com}}


%


\maketitle

\begin{abstract}
We consider quantum MDS (QMDS) codes for quantum systems
of dimension $q$ with lengths up to $q^2+2$ and minimum distances up to
$q+1$. We show how starting from QMDS codes of length $q^2+1$
based on  cyclic and constacyclic codes, new QMDS codes can be
obtained by shortening.  
We provide numerical evidence for our conjecture that almost all
admissible lengths, from a lower bound $n_0(q,d)$ on, are achievable by
shortening. Some additional codes that fill gaps in the list of
achievable lengths are presented as well along with a construction of
a family of QMDS codes of length $q^2+2$, where $q=2^m$, that
appears to be new.
\end{abstract}


%
\IEEEpeerreviewmaketitle


\section{Introduction}
\IEEEPARstart{Q}{uantum} error-correcting codes (QECC) are a key ingredient to
implement information processing based on quantum mechanics. For
quantum systems composed of $n$ subsystems of dimension $q{\geq}2$, so-called
\emph{qudits}, a quantum code $\mathcal{C}=(\!(N,K)\!)_q$ is a
$k$-dimensional subspace of the Hilbert space $(\C^q)^{\otimes n}$. If
the dimension of the code $\mathcal{C}$ is $q^k$, it will be denoted
by $\mathcal{C}=[\![n,k,d]\!]_q$, where $d$ is the minimum distance. A
code with minimum distance $d$ is able to correct errors that
affect no more than $(d-1)/2$ of the subsystems.  The quantum
Singleton bound \cite{KnLa97,Rai99} relates the parameters $n$, $k$,
and $d$ as follows:
\begin{equation}\label{eq:quantum_Singleton}
n+2\ge k+2d
\end{equation}
A quantum code for which equality holds in
(\ref{eq:quantum_Singleton}) is called a quantum MDS (QMDS) code.  For
classical codes, the existence of an MDS code $C=[n,k,n+1-k]_q$
implies the existence of MDS codes $C'=[n',k',n'+1-k']_q$ for all $k'\le k$,
$n'\le n$, $k\le n'$. For quantum codes, this is not true in general,
i.\,e., a QMDS code $\mathcal{C}=[\![n,n+2-2d,d]\!]_q$ does not
necessarily imply the existence of QMDS codes of smaller length or
smaller dimension.

For any number of qudits, the full space is a trivial QMDS code
$\mathcal{C}=[\![n,n,1]\!]_q$, where $q>1$ can be any integer, not
necessarily a prime power.  QMDS codes with distance $d=2$ exist for
even length $n$ when $q=2$ (see \cite{Rai99:dist2}), and for all
lengths $n\ge 2$ when $q>2$ is a prime power (see below). This implies
the existence of QMDS codes with $d=2$ for all lengths $n\ge 2$ when
$q$ is odd or divisible by $4$.

When the length of the code is bound by $n\le q$, QMDS codes can be
obtained from extended Reed-Solomon codes (see, e.g., \cite{GBR04}).
Single-error-correcting QMDS codes for length $4\le n\le q^2+1$ have
been discussed in \cite{LiXu10} for odd prime powers $q$, and more
generally in \cite{Gra10,JLLX10}.

Quantum MDS codes of length $n\in\{q^2-1,q^2,q^2+1\}$ have been
discussed in \cite{RGB04,JLLX10}. In \cite{JLLX10} there are also QMDS
codes of length $n$ in the range $q{+}1<n< q^2{-}1$, with the minimum
distance $d$ bounded by $d\le (q+5)/4$.  In \cite{SaKl05}, QMDS codes
for certain lengths in the range $q{+}1<n<q^2{-}1$ were constructed
based on generalized Reed-Muller codes.

More recently, QMDS codes with a larger range for the minimum distance
based on cyclic and constacyclic codes have been derived (see, e.g.,
\cite{KaZh13,ZhCh14,WaZh14,CLZ14}).  Those constructions put some
constraints on the length $n$ of the code, e.g., $n$ has to be a
divisor of $q^2\pm 1$, and in most cases, the minimum distance is
bounded by some fraction of $q$.

Here, extending our results from \cite{RGB04}, we show that QMDS codes
exists for essentially all lengths $n$ in the range $n_0(q,d)\le
q^2+1$, where the lower bound $n_0$ grows with the minimum distance
$d$.  For most of these QMDS codes, the minimum distance is bounded by
$d\le q+1$, but we present also some examples of qutrit and ququad
QMDS codes exceeding this bound.

After recalling basic results about stabilizer codes and construction of classical MDS codes in Section
\ref{sec:MDS}, Section \ref{sec:PC} presents the main technique how
shorter QMDS codes with the same minimum distance can be
obtained. Theoretical results are summarized in Section
\ref{sec:results}, supplemented by computational results in Section
\ref{sec:comp_results}.  All computations have been performed using
the computer algebra system Magma \cite{magma}.

\section{Stabilizer Codes}\label{sec:QECC}
Most quantum error-correcting codes are so-called stabilizer codes.
Here we briefly summarize the basic results which are relevant in our
context (for more details, see e.g. \cite{CRSS98,AsKn01,KKKS06}).

The construction of stabilizer codes is based on classical codes which
are self-orthogonal with respect to a symplectic inner product.  The
most general construction of a stabilizer code for qudits starts with
an additive code $C=(n,p^\ell)_{q^2}$ of length $n$ over the a
quadratic extension field $\F_{q^2}$.  Note that the code does not
need to be $\F_q$- or $\F_{q^2}$-linear, but just $\F_p$-linear, where
$q=p^m$ and $p$ is prime.  

Here we consider only the special case that the code is
$\F_{q^2}$-linear.  In this case, the symplectic inner product is
equivalent to the so-called Hermitian inner product. For vectors
$\bm{c},\bm{c}'\in\F_{q^2}^n$, it is defined as
\begin{alignat}{5}
\bm{c}*\bm{c}'=\sum_{i=1}^n c_i^q c_i'.
\end{alignat}
We consider the dual code with respect to this Hermitian inner product.
\begin{definition}[Hermitian dual code]
Given a linear code $C=[n,k]_{q^2}$ over $\F_{q^2}$, the
Hermitian dual $C^*$ is given by
\begin{alignat}{5}
C^*=\{ \bm{v}\colon\bm{v}\in\F_{q^2}^n\mid \forall\bm{c}\in C\colon \bm{c}*\bm{v}=0\}.
\end{alignat}
The Hermitian dual code $C^*=[n,n-k]_{q^2}$ is an $\F_{q^2}$-linear
code of dimension $n-k$.
\end{definition}
The following proposition is a central result relating classical
codes and quantum stabilizer codes (see, e.g., \cite[Corollary
  19]{KKKS06}).
\begin{proposition}
Let $C=[n,n-k]_{q^2}$ be an $\F_{q^2}$-linear code that is contained
in its Hermitian dual $C^*=[n,k,d^*]_{q^2}$. Then there exists a
quantum stabilizer code $\mathcal{C}=[\![n,2k-n,d]\!]_q$.  The minimum
distance $d$ is given by
\begin{alignat}{5}\label{eq:min_dist}
d=\min\{\wgt\bm{c}\colon\bm{c}\in C^*\setminus C\}\ge d^*.
\end{alignat}
If equality holds in (\ref{eq:min_dist}), the code is said to be
\emph{pure}.
\end{proposition}
In \cite{Rai99} it has been shown that a QMDS code is always pure.

Shortening of the self-orthogonal code $C$ yields the
following derivation rule (see also \cite[Lemma 70]{KKKS06}).
\begin{proposition}\label{prop:puncturing}
Assume that there is a pure stabilizer code
$\mathcal{C}=[\![n,k,d]\!]_q$ with $d>1$. Then there exists a QECC
$\mathcal{C}'=[\![n-1,k+1,d-1]\!]_q$.
\end{proposition}
\begin{IEEEproof}(sketch)
When we puncture the code $C^{*}$ corresponding to $\mathcal{C}$ at
say the first position, we obtain a code ${C'}^*$ of length $n-1$
which has the same number of codewords as $C^*$ and minimum distance
$d'\ge d-1$.  The code $({C'}^*)^*=C'$ contains all vectors $\bm{c}'$
for which $0\bm{c}'\in C$.  Hence $C'\subset {C'}^*$. The dimension of
$C'$ is one less than the dimension of $C$, resulting in an increase of
the dimension of the quantum code by one.
\end{IEEEproof}
Note that shortening of the code $C^*$ corresponds to puncturing the
code $C$. However, after puncturing, the code $C'$ need no longer be
self-orthogonal with respect to the symplectic inner product.

Repeated application of Proposition \ref{prop:puncturing} yields the
following.
\begin{corollary}\label{corr:shorten_stabilizer}
Assume that a QMDS code $\mathcal{C}=[\![n,n+2-2d,d]\!]_q$
exists. Then for all $0\le s<d$, there exist also QMDS codes
$\mathcal{C}'=[\![n-s,n+s+2-2d,d-s]\!]_q$.
\end{corollary}

\section{Classical MDS Codes}\label{sec:MDS}
In order to construct quantum MDS codes of length $q^2+1$, we start
with cyclic or constacyclic MDS codes (see also \cite{RGB04,GrGu08}).
In order to simplify the notation, without loss of generality, we
consider codes over the field $\F_q$ instead of the field $\F_{q^2}$.
\begin{theorem}\label{theorem:cyclicMDS}
For any $k$, $1\le k\le q+1$, there exists a $[q+1,k,q-k+2]_q$ MDS code
over $\F_q$ that is either cyclic or constacyclic.
\end{theorem}
\begin{IEEEproof}
Let $\omega$ denote a primitive element of $\F_{q^2}$. Hence
$\alpha:=\omega^{q-1}$ is a primitive $(q+1)$-th root of unity.
\nopagebreak

First we consider the case when $q+1-k$ is odd. We define the
following polynomial of degree $2\mu+1$:
\begin{equation}\label{eq:odd_deg_MDSpoly}
g_1(z):=\prod_{i=-\mu}^\mu (z-\alpha^i).
\end{equation}
Its zeros $\alpha^i$ and $\alpha^{-i}$ are conjugates of each other
since $\alpha^q=\alpha^{-1}$. Hence $g_1(z)$ a polynomial over
$\F_q$. The resulting cyclic code $C$ over $\F_q$ has length $q+1$ and
dimension $q-2\mu$. The generator polynomial $g_1(z)$ has $2\mu+1$
consecutive zeros, so the BCH bound yields $d\ge 2\mu+2$. Therefore
$C$ is an MDS code $[q+1,q-2\mu,2\mu+2]_q$.

If $q+1-k$ is even and $q$ is even too, the polynomial
\begin{alignat}{3}
g_2(z)&:=\prod_{i=q/2-\mu}^{q/2+1+\mu} (z-\alpha^i)\nonumber\\
&=\prod_{i=q/2-\mu}^{q/2} (z-\alpha^i)(z-\alpha^{-i})
\end{alignat}
has degree $2\mu+2$. It is a polynomial over $\F_q$ with $2\mu+2$
consecutive zeros, so the resulting code is an MDS code with parameters
$[q+1,q-1-2\mu,2\mu+3]_q$.

Finally, if $q+1-k$ is even and $q$ is odd, consider the polynomial
\begin{alignat}{3}
g_3(z):=\prod_{i=1}^\mu(z-\omega \alpha^i)(z-\omega \alpha^{1-i})\label{eq:gen_poly3}
\end{alignat}
of degree $2\mu$. The roots $\omega \alpha^i$ and $\omega
\alpha^{1-i}$ are conjugates of each other as
$(\omega\alpha^i)^q=\omega^{(1+(q-1)i)q}
=\omega^{q+(1-q)i}=\omega^{1+(q-1)(1-i)}=\omega\alpha^{1-i}$, so
$g_3(z)$ is a polynomial over $\F_q$. Furthermore, $g_3(z)$ divides
$z^{q+1}-\omega^{q+1}\in\F_{q^2}[z]$ as
$(\omega\alpha^i)^{q+1}=\omega^{q+1}$.  Therefore $g_3(z)$ defines a
constacyclic code $C$ of length $q+1$ and dimension $q+1-2\mu$ over
$\F_q$. From the analogue of the BCH bound for constacyclic codes
(see, e.g., \cite{RvZ08}), we have $d\ge 2\mu+1$. Hence $C$ is an MDS code with
parameters $[q+1,q+1-2\mu,2\mu+1]_q$.
\end{IEEEproof}
\begin{remark}
Theorem \ref{theorem:cyclicMDS} is a slightly modified version of
Theorem 9 in \cite[Ch.~11, \S 5]{MS77}.  There only cyclic codes are
considered; the construction fails when both $q$ and $k$ are odd (see
also the preface to the third printing of \cite{MS77}).
\end{remark}

\section{Shortening Quantum Codes}\label{sec:PC}
While classical linear codes can be shortened to any length, i.e.,
from a code $[n,k,d]$ one obtains a code $[n-r,k'\ge k-r,d'\ge d]$ for
any $r$, $0\le r\le k$, this is in general not true for quantum
codes. However, in \cite{Rai99} it is shown how quantum codes can be
shortened using the so-called puncture code.  Here we recall the main
results for $\F_{q^2}$-linear codes.
\begin{definition}[puncture code]
Let $C=[n,k]_{q^2}$ be an $\F_{q^2}$-linear code.
The puncture code of $C$ is defined as
\begin{equation}\label{eq:PC_linear}
P(C):=\Bigl\langle \{ (c_i^q c_i')_{i=1}^n\colon \bm{c},\bm{c}'\in
C\Bigr\rangle^\bot\cap \F_q^n,
\end{equation}
where the angle brackets denote the $\F_{q^2}$-linear span.
\end{definition}

From \cite[Theorem~3]{Rai99} we get:
\begin{theorem}\label{theorem:puncture}
Let $C=[n,k]_{q^2}$ be an $\F_{q^2}$-linear code, not necessarily
self-orthogonal, of length $n$ and dimension $k$ such that the
Hermitian dual code $C^*=[n,n-k]_{q^{2}}$ has minimum distance $d$. If
there exists a codeword in $P(C)$ of weight $r$, then there exists a
pure QECC $\QECC(r,k',d',q)$ for some $k'\ge r-2k$ and $d'\ge d$.
\end{theorem}
\begin{IEEEproof}
Let $\bm{x}\in P(C)$ be a codeword of weight $r$ and let $S=\{i\colon
i\in \{1,\ldots,n\}\mid x_i\ne 0\}$ denote its support. Note that the
norm $\mathcal{N}_{\F_{q^2}/\F_q}(x)=x^{q+1}$ is surjective.  Hence
there exists a vector $\bm{y}\in\F_{q^2}^n$ such that
$y_i^{q+1}=x_i$ for $1\le i\le n$.
We define the code $\widetilde{C}$ to be
\begin{equation}\label{eq:C_tilde}
\widetilde{C}:=\Bigl\{ (y_i c_i)_{i=1}^n\colon \bm{c}\in C\Bigr\},
\end{equation}
i,\,e., we pointwise multiply the codewords by the corresponding
elements of $\bm{y}$. For arbitrary
$\widetilde{\bm{c}},\widetilde{\bm{c}}'\in\widetilde{C}$, we get
\begin{alignat}{5}
\widetilde{\bm{c}}*\widetilde{\bm{c}}'
&=\sum_{i=1}^n  \widetilde{c}_i^q \widetilde{c}_i'
&=\sum_{i=1}^n  (y_ic_i)^q y_i c_i'
&=\sum_{i=1}^n  x_i c_i^q  c_i'.\label{eq:proof_PC}
\end{alignat}
From (\ref{eq:PC_linear}) it follows that (\ref{eq:proof_PC}) vanishes,
i.\,e., $\widetilde{C}$ is self-orthogonal. As (\ref{eq:proof_PC})
depends only on the coordinates of $\bm{x}$ that are non-zero, we can
delete the other positions in $\widetilde{C}$ and obtain an
$\F_{q^2}$-linear self-orthogonal code $D\subseteq \F_{q^2}^r$ given
by
\begin{alignat*}{5}
D:=\Bigr\{ (y_i c_i)_{i\in S}\colon \bm{c}\in C\Bigr\}.
\end{alignat*}
Puncturing the code $\widetilde{C}$ may reduce its dimension. Hence
$D$ has parameters $D=[n,\widetilde{k}]_{q^2}$ for some
$\widetilde{k}\le k$. The dual code $D^*$ is obtained by shortening
the code $C^*$, and multiplying the resulting codewords by the
corresponding non-zero entries of $\bm{y}$.  Hence the minimum distance
$d'$ of $D^*$ is not smaller than the minimum distance of $C^*$. This
shows $d'\ge d$. Overall, we get a quantum code with parameters
$\mathcal{C}'=[\![r,k',d']\!]_q$, where $k'=r-2\widetilde{k}\ge r-2k$.
\end{IEEEproof}
It should be stressed that the puncture code $P(C)$ can be computed
for any code $C$, not only for self-orthogonal ones.  In particular,
using a codeword of maximal weight in $P(C)$, an arbitrary linear code
can be converted into a self-orthogonal code.

The following obvious lemma will prove useful:
\begin{lemma}\label{lemma:nested_PC}
If $C_1\subseteq C_2$, then $P(C_2)\subseteq P(C_1)$.
\end{lemma}

For cyclic or constacyclic linear codes over $\F_{q^2}$, the puncture
code will be again cyclic or constacyclic, respectively.  We have the
following characterization:
\begin{theorem}
Let $C^*=[n,k]_{q^2}$ be an $\F_{q^2}$-linear cyclic code with
defining set $\mathcal{Z}$, i.e., the generator polynomial $g(x)$ of
$C^{*}$ has roots $\{\alpha^i\colon i\in\mathcal{Z}\}$ where $\alpha$
is a primitive $n$-th root of unity.  For a constacyclic code
$C^*=[n,k]_{q^2}$ with shift constant $\beta^n$, the generator
polynomial $g(x)$ is a divisor of $x^n-\beta^n$, and its roots can be
expressed as $\{\beta\alpha^i\colon i\in\mathcal{Z}\}$.

Then the puncture code $P(C)$ is a (consta)cyclic code over $\F_q$
with defining set
\begin{alignat}{5}
\mathcal{Z}'=\{iq+jq^2\colon i,j \in \mathcal{Z}\}.\label{eq:roots_PC}
\end{alignat}
\end{theorem}
\begin{IEEEproof}
First note that a cyclic code is a constacyclic code with shift
constant $\beta^n=1$.  A parity check matrix for a (consta)cyclic code
is given by
\begin{alignat}{5}\label{eq:check1}
H=\biggl((\beta\alpha^i)^0,(\beta\alpha^i)^1,\ldots,(\beta\alpha^i)^{n-1}\biggr)_{i\in\mathcal{Z}}.
\end{alignat}
For an $\F_{q^2}$-linear code, the symplectic dual code equals the
Hermitian dual code, which is the code obtained by Galois conjugation
of the usual dual code. Therefore, a generator matrix of $C$ is given
by
\begin{alignat}{5}
G&=\biggl((\beta\alpha^i)^0,(\beta\alpha^i)^q,\ldots,(\beta\alpha^i)^{(n-1)q}\biggr)_{i\in\mathcal{Z}}\nonumber\\
 &=\biggl((\beta^q\alpha^i)^0,(\beta^q\alpha^i),\ldots,(\beta^q\alpha^i)^{{n-1}}\biggr)_{i\in\mathcal{Z}^q},
\end{alignat}
where $\mathcal{Z}^q=\{i^q\colon i\in\mathcal{Z}\}$. From
(\ref{eq:PC_linear}) it follows that a parity check matrix of $P(C)$
is given by the component-wise product of the rows of $G$ and their
Galois conjugates:
\begin{alignat}{5}
&H_{P(C)}\nonumber\\
&\quad=\biggl(
(\beta^q\alpha^i)^0(\beta^{q^2}\alpha^{qj})^0,
\ldots
(\beta^q\alpha^i)^{n-1}(\beta^{q^2}\alpha^{qj})^{n-1}
\biggr)_{i,j\in\mathcal{Z}^q}\nonumber\\
&\quad=\biggl(
(\tilde{\beta}\alpha^{i+qj})^0,
(\tilde{\beta}\alpha^{i+qj}),
\ldots
(\tilde{\beta}\alpha^{i+qj})^{n-1}
\biggr)_{i,j\in\mathcal{Z}^q},
\end{alignat}
where $\tilde{\beta}=\beta^{q(q+1)}$. Note that $\beta^n\in\F_{q^2}$,
and hence $\tilde{\beta}^n=(\beta^n)^{q(q+1)}=(\beta^n)^{q+1}\in\F_q$,
as for an element $x\in\F_{q^2}$, its norm
$x^{q+1}\in\F_q$. Therefore, $P(C)$ is a constacyclic code with shift
constant $\tilde{\beta}^n$, and its generator polynomial has roots
$\{\tilde{\beta}\alpha^i\colon i,j\in\mathcal{Z}'\}$, where
$\mathcal{Z}'$ is defined in (\ref{eq:roots_PC}).
\end{IEEEproof}
For the MDS codes from Theorem \ref{theorem:cyclicMDS}, the defining
set $\mathcal{Z}$ consists of $d-1$ consecutive numbers.  Based on the
computational results in Section \ref{sec:comp_results} below, we have
the following conjecture for the corresponding puncture code:
\begin{conjecture}\label{conj:PC_dmin}
Let $C^*=[q^2+1,q^2=1-d,d]_{q^2}$ be an $\F_{q^2}$-linear
(consta)cyclic MDS code.

Then the corresponding puncture code $P(C)$ has parameters
$PC=[q^2+1,q^2+1-(d-1)^2,d']_q$ where
\begin{alignat}{5}
d'=
\left\{
\begin{array}{@{}l@{\;}l@{}}
2(d-1) & \text{for $1< d \le q/2+1$}\\
(q+1)(d-1-\lfloor q/2\rfloor)) & \text{for $q/2+1< d\le q$, $q$ odd}\\
q(d-\lfloor q/2\rfloor) & \text{for $q/2+1< d \le q$, $q$ even}\\
q^2+1 & \text{for $d=q+1$}
\end{array}
\right.
\end{alignat}
\end{conjecture}
Using the Hartmann-Tzeng bound for (consta)cyclic codes (see, e.g.,
\cite{RvZ08}), we get the lower bound $d'\ge 2(d-1)$. Using the Roos
bound, we get a better lower bound for $d>q/2+1$, but in general the
conjectured minimum distance $d'$ is even larger.

\section{Results}\label{sec:results}
\subsection{QMDS Codes of Minimum Distance Two}
In \cite[Table II]{KKKS06}, the existence of QMDS codes
$\mathcal{C}=[\![n,n-2,2]\!]_{q}$ is stated for the case that the
length $n$ is divisible by the characteristic $p$ of the field $\F_q$,
i.e., $q=p^m$, $p$ prime.  In that case, the classical repetition code
$C=[n,1,n]_q$ is contained in its Euclidian dual $C^\bot=[n,n-1,2]_q$,
and the CSS construction yields the corresponding quantum code.  More
generally, consider the classical repetition code $C=[n,1,n]_{q^2}$
over the field $\F_{q^2}$.  Then the puncture code is the MDS code
$P(C)=[n,n-1,2]_q$.  When $q>2$, $P(C)$ contains words of all weights
$2\le w\le n$  (see \cite{EGS11}).  
By \cite[Theorem 14]{Rai99}, the existence of codes
$[\![n,n-2,2]\!]_{q_1}$ and $[\![n,n-2,2]\!]_{q_2}$ implies the
existence of a code $[\![n,n-2,2]\!]_{q_1 q_2}$.  In summary, we have
\begin{theorem}
Let $q>1$ be an arbitrary integer, not necessarily a prime power.
Quantum MDS codes $\mathcal{C}=[\![n,n-2,2]\!]_{q}$ exist for all even
length $n$, and for all length $n\ge 2$ when the dimension $q$ of the
quantum systems is an odd integer or is divisible by $4$.
\end{theorem}

\subsection{QMDS Codes of Length $q^2+1$}
\begin{theorem}
Our construction yields QMDS codes with parameters
$\mathcal{C}=[\![q^2+1,q^2+3-2d,d]\!]_q$ for all $1\le d\le q+1$ when
$q$ is odd, or when $q$ is even and $d$ is odd.
\end{theorem}
\begin{IEEEproof}
When the minimum distance is $d=q+1$, the puncture code has $q^2$
consecutive roots and is hence a trivial MDS code
$P(C)=[q^2+1,1,q^2+1]_q$.  In particular, $P(C)$ contains a word of
weight $q^2+1$, and hence a QMDS code
$\mathcal{C}=[\![q^2+1,q^2+1-2q,q+1]\!]_q$ exists.
Note that the MDS codes of even dimension from Theorem
\ref{theorem:cyclicMDS} form a chain of nested codes, and likewise the
codes of odd dimension.  Using Lemma \ref{lemma:nested_PC}, it follows
that $P(C)$ contains a word of weight $q^2+1$ whenever the minimum
distance $d$ of $C^*$ has the same parity as $q+1$.

When $q$ is odd and $d=q$, the classical MDS code is a constacyclic
code with generator polynomial $g_3(z)$ given by (\ref{eq:gen_poly3}).
The corresponding puncture code $P(C)$ is a constacyclic code over
$\F_q$ of length $q^2+1$ with defining set $\mathcal{Z}'=\{i+qj\colon
i,j=-t+1,\ldots,t\}$, $t=(q-1)/2$.  The generator polynomial of $P(C)$
divides the polynomial $f(z)=z^{q^2+1}-\gamma$, where $\gamma\in\F_q$
is a primitive element of $\F_q$. As $q$ is odd, $\gamma$ has two
square roots $\pm\sqrt{\gamma}\in\F_{q^2}$.  Moreover, for any
$x\in\F_{q^2}$, $x^{q^2+1}=x^2$ which shows that $\pm\sqrt{\gamma}$
are roots of $f(z)$ and hence the polynomial $f(z)$ is divisible by
$z^2-\gamma$. The corresponding defining set is
$\{(q^2+1)/2,(q^4+q^2)/2+1\}$ which is disjoint from $\mathcal{Z}'$.
Hence $\pm\gamma$ are not among the roots of $g_3(z)$ and $g_3(z)$
divides the polynomial $g(z)=f(z)/(z^2-\gamma)$.  The constacyclic
code generated by $g(z)$ is the sum of two trivial MDS codes
$[(q^2+1)/2,1,(q^2+1)/2]_q$.  In particular it contains a word of
weight $q^2+1$.  Using Lemma \ref{lemma:nested_PC}, it follows that
$P(C)$ contains a word of weight $q^2+1$ for $q$ odd and when the
minimum distance $d$ of $C^*$ is odd as well.
\end{IEEEproof}
Note that when both $q$ and $d$ are even, the theorem does not hold in
general. For example, our construction does not yield a QMDS code
$[\![17,11,4]]_4$.  However, this might be the only exception as for
$q=2^m$, $m=3,4,5,6,7$, our construction provides codes
$[\![4^m+1,4^m+3-2^{m+1},2^m]\!]_{2^m}$.  To show this, we find a
cyclic subcode of $P(C)$ of dimension $4$ which contains a word of
weight $4^m+1$. This also implies the existence of QMDS codes of
length $2^m+1$ for all distances $d\le 2^m+1$.

\subsection{QMDS Codes of Length $q^2+2$}
For $q^2=2^{2m}$, there exist classical MDS codes with parameters
$C^*=[2^{2m}+2,2^{2m}-1,4]_{2^{2m}}$ (see, e.g., \cite[Ch.~11, \S 5, Theorem
  10]{MS77}). A parity check matrix for $C^*$ is 
\begin{alignat}{5}\label{eq:parity_check_char_two}
H=\left(
\begin{array}{cccccccccccccccccc}
1&1&1&\ldots&1 & 1 & 0 & 0\\
\alpha^0 &\alpha^1 &\alpha^2 &\ldots&\alpha^{q^2-2} & 0 & 1 & 0\\
\alpha^0 &\alpha^2 &\alpha^4 &\ldots&\alpha^{2(q^2-2)} & 0 & 0 & 1
\end{array}
\right),
\end{alignat}
where $\alpha$ denotes a primitive element of $\F_{q^2}$.  
The code $C$ is not self-orthogonal, as, for example, the Hermitian
inner product of the second row of $H$ with itself is non-zero.
\begin{theorem}
For $q=2^m$, there exist QMDS codes with parameters
$\mathcal{C}=[\![4^m+2,4^m-4,4]\!]_{2^m}$.
\end{theorem}
\begin{IEEEproof}
A parity check matrix $H_{P(C)}$ of the puncture code $P(C)$ is given
by
\begin{alignat}{5}\label{eq:parity_check_PC_char_two}
\left(
\begin{array}{cccccccccccccccccc}
1&1&1&\ldots&1 & 1 & 0 & 0\\
1 &\alpha^{q+1} &\alpha^{2(q+1)} &\ldots&\alpha^{(q^2-2)(q+1)} & 0 & 1 & 0\\
1 &\alpha^{2(q+1)} &\alpha^{4(q+1)} &\ldots&\alpha^{2(q^2-2)(q+1)} & 0 & 0 & 1\\
1 &\alpha^1 &\alpha^2 &\ldots&\alpha^{q^2-2} & 0 & 0 & 0\\
1 &\alpha^2 &\alpha^4 &\ldots&\alpha^{2(q^2-2)} & 0 & 0 & 0\\
1 &\alpha^q &\alpha^{2q} &\ldots&\alpha^{q(q^2-2)} & 0 & 0 & 0\\
1 &\alpha^{q+2} &\alpha^{2(q+2)} &\ldots&\alpha^{(q+2)(q^2-2)} & 0 & 0 & 0\\
1 &\alpha^{2q} &\alpha^{4q} &\ldots&\alpha^{2q(q^2-2)} & 0 & 0 & 0\\
1 &\alpha^{2q+1} &\alpha^{4q+2} &\ldots&\alpha^{(2q+1)(q^2-2)} & 0 & 0 & 0
\end{array}
\right),
\end{alignat}
where $\alpha$ denotes a primitive element of $\F_{q^2}$.  
The first three rows of $H_{P(C)}$ are given by the componentwise norm
of the matrix $H$ defined in
(\ref{eq:parity_check_char_two}). In particular, the entries of these
three rows lie in $\F_q$.  Consider the matrix
\begin{alignat}{5}
G_1=&\left(\arraycolsep0.5\arraycolsep
\begin{array}{cccccccccccccccccc}
1&1&1&\ldots&1 & 1 & 0 & 0\\
1 &\alpha^{-(q+1)} &\alpha^{-2(q+1)} &\ldots&\alpha^{-(q^2-2)(q+1)} & 0 & 1 & 0\\
1 &\alpha^{-2(q+1)} &\alpha^{-4(q+1)} &\ldots&\alpha^{-2(q^2-2)(q+1)} & 0 & 0 & 1
\end{array}
\right)\nonumber\\
=&\left(
\begin{array}{cccccccccccccccccc}
1&1&1&\ldots&1 & 1 & 0 & 0\\
1 &\beta &\beta^2 &\ldots&\beta^{q^2-2} & 0 & 1 & 0\\
1 &\beta^2 &\beta^4 &\ldots&\beta^{2(q^2-2)} & 0 & 0 & 1
\end{array}
\right),
\end{alignat}
where $\beta=\alpha^{-(q+1)}$ is a primitive element of $\F_q$.  The
matrix $G_1$ is obtained by taking the inverse of the non-zero entries
in the first three rows of $H_{P(C)}$ given in
(\ref{eq:parity_check_PC_char_two}).  Each row of $G_1$ is orthogonal
to all rows of $H_{P(C)}$, i.e., the span of $G_1$ is contained in
$P(C)$.  Let $f(x)=x^2+\gamma_1 x+\gamma_0$ be an irreducible
polynomial over $\F_q$. Then in particular $\gamma_1\ne 0\ne\gamma_0$.
Furthermore, every coordinate of the codeword
$\bm{c}=(\gamma_0,\gamma_1,1)G_1$ is non-zero, as the last three
coordinates of $\bm{c}$ just equal $(\gamma_0,\gamma_1,1)$, and the
other coordinates correspond to the evaluation of the irreducible
polynomial $f(x)$ at some power of $\beta$.  Therefore, $P(C)$
contains the codeword $\bm{c}$ of weight $q^2+2$.
\end{IEEEproof}
The computational results show that for $q=2^m$, $m=3,\ldots,7$ the
puncture code $P(C)$ of the MDS code with parity check matrix $H$
given in (\ref{eq:parity_check_char_two}) does not only contain a word
of weight $4^m+2$, but words of all weights $6\le w\le 4^m+2$. This
suggests the following:
\begin{conjecture}
For $q=2^m$, $q\ne 4$, there exist QMDS codes with parameters
$[\![n,n-6,4]\!]_{2^m}$ for all $6\le n\le 4^m+2=q^2+2$.
\end{conjecture}

\section{Computational Results}\label{sec:comp_results}
\subsection{Qubit Codes}
For $q=2$, we only have the QMDS codes $[\![5,1,3]\!]_2$,
$[\![6,0,4]\!]_2$, as well as the codes of even length
$[\![2m,2m-2,2]\!]_2$ and the trivial codes $[\![n,n,1]\!]_2$.

\subsection{Qutrit Codes}
For $q=3$, in addition to the trivial codes with distance $d=1,2$, our
construction yields QMDS codes $[\![n,n-4,3]\!]_3$ for $n=4,\ldots,10$
as well as the code $[\![10,4,4]\!]_3$.

What is more, from Hermitian self-dual MDS codes $C=[6,3,4]_9$ and
$C=[10,5,6]_9$ in \cite{GrGu09}, we get QMDS codes
$\mathcal{C}=[\![6,0,4]\!]_3$ and $\mathcal{C}=[\![10,0,6]\!]_3$,
respectively.  By Corollary \ref{corr:shorten_stabilizer}, from
$\mathcal{C}$ we obtain QMDS codes $[\![9,1,5]\!]_3$ and
$[\![8,2,4]\!]_3$

\subsection{Ququad Codes}
For $q=4$, in addition to the trivial codes with distance $d=1,2$, our
construction yields QMDS codes $[\![n,n-4,3]\!]_4$ for
$n=4,\ldots,17$, codes $[\![n,n-6,4]\!]_4$ for even
$n=8,10,\ldots,18$, as well the code $[\![17,9,5]\!]_4$.  Codes with
distance $d=4$ and odd length cannot be directly obtained, as the
puncture code is even in this case. However, by direct search we found
codes with parameters $[\![6,0,4]\!]_4$, $[\![9,3,4]\!]_4$, and
$[\![11,5,4]\!]_4$.  From a Hermitian self-dual MDS code of length
$10$ given in \cite{GrGu08}, we obtain QMDS codes $[\![10,0,6]\!]_4$
and $[\![9,1,5]\!]_4$.

\subsection{Ququint Codes}
For $q=5$, in addition to the trivial codes with distance $d=1,2$, our
construction yields QMDS codes $[\![n,n-4,3]\!]_5$ for
$n=4,\ldots,26$, codes $[\![n,n-6,4]\!]_5$ for $n=6$ and
$n=8,\ldots,18$, codes $[\![n,n-8,5]\!]_5$ for $n=12,\ldots,26$, and
the code $[\![26,16,6]\!]_5$.  For $d=4$, the puncture code does not
contain a word of weight $7$.  Hermitian self-dual MDS codes from
\cite{GrGu08} yield QMDS codes $[\![8,0,5]\!]_5$ and
$[\![10,0,6]\!]_5$.  Applying Corollary \ref{corr:shorten_stabilizer}
to these codes, we obtain the missing QMDS code $[\![7,1,4]\!]_5$ and
in addition a code $[\![9,1,5]\!]_5$.  Moreover, a QMDS code
$[\![10,2,5]\!]_5$ was found by randomized search.

\subsection{Qusept Codes}
For $q=7$, we obtain all QMDS codes $[\![n,n+2-2d,d]\!]_7$ for all
$2d-2\le n\le 50$, $2\le d\le 4$.  For $d=5$, the puncture codes does
not contain words of weight $w=9,10,11$, and for $d=6$, $P(C)$
contains words in the range $16,\ldots,50$, with the exception of
$w=17$. For $d=7$, $P(C)$ contains words in the range $16,\ldots,50$,
with the exception of $w=26,27,29$.  For $d=8$ we have the code
$[\![50,36,8]\!]_7$ from our construction.  From Hermitian self-dual
MDS codes in \cite{GrGu08}, we obtain QMDS codes $[\![10,0,6]\!]_7$,
$[\![12,0,7]\!]_7$, and $[\![14,0,8]\!]_7$.  Using Corollary
\ref{corr:shorten_stabilizer}, we also get the missing codes
$[\![9,1,5]\!]_7$, $[\![10,2,5]\!]_7$, and $[\![11,3,5]\!]_7$ with
distance $5$, as well as the additional shorter codes
$[\![11,1,6]\!]_7$, $[\![13,1,7]\!]_7$, and $[\![12,2,6]\!]_7$.  We
have not yet found a code $[\![17,7,6]\!]_7$ or the QMDS codes of
length $n=26,27,29$ and distance $7$.

\subsection{Quoct Codes}
For $q=8$, we obtain almost all of the QMDS codes implied by
Conjecture \ref{conj:PC_dmin}.  For $d=8$, the puncture code $P(C)$
does not contain words of weight $33,34,35,37,39$.  For $d=7$, we have
all weights in the range $24,\ldots,65$.  For $d=6$, the puncture code
$P(C)$ contains words of weights $16,18,\ldots,65$, but random
sampling did not reveal a word of weight $17$. The corresponding code
$[\![17,5,6]\!]$ has not yet been found either.  For $d=5$, random
sampling of the puncture code $P(C)$ did not reveal words of weight
$w=9,10,11$, but the corresponding codes $[\![9,1,5]\!]_8$,
$[\![10,2,5]\!]_8$, and $[\![11,3,5]\!]_8$ can be derived from
Hermitian self-dual codes $[10,5,6]_{64}$, $[12,6,7]_{64}$, and
$[14,7,8]_{64}$.

\subsection{Qudit Codes with $9\le q\le 32$, $q=64$}
\begin{figure}[!ht]
\centering\scriptsize
\begin{tabular}{@{}cc@{}}
\includegraphics[width=0.47\hsize]{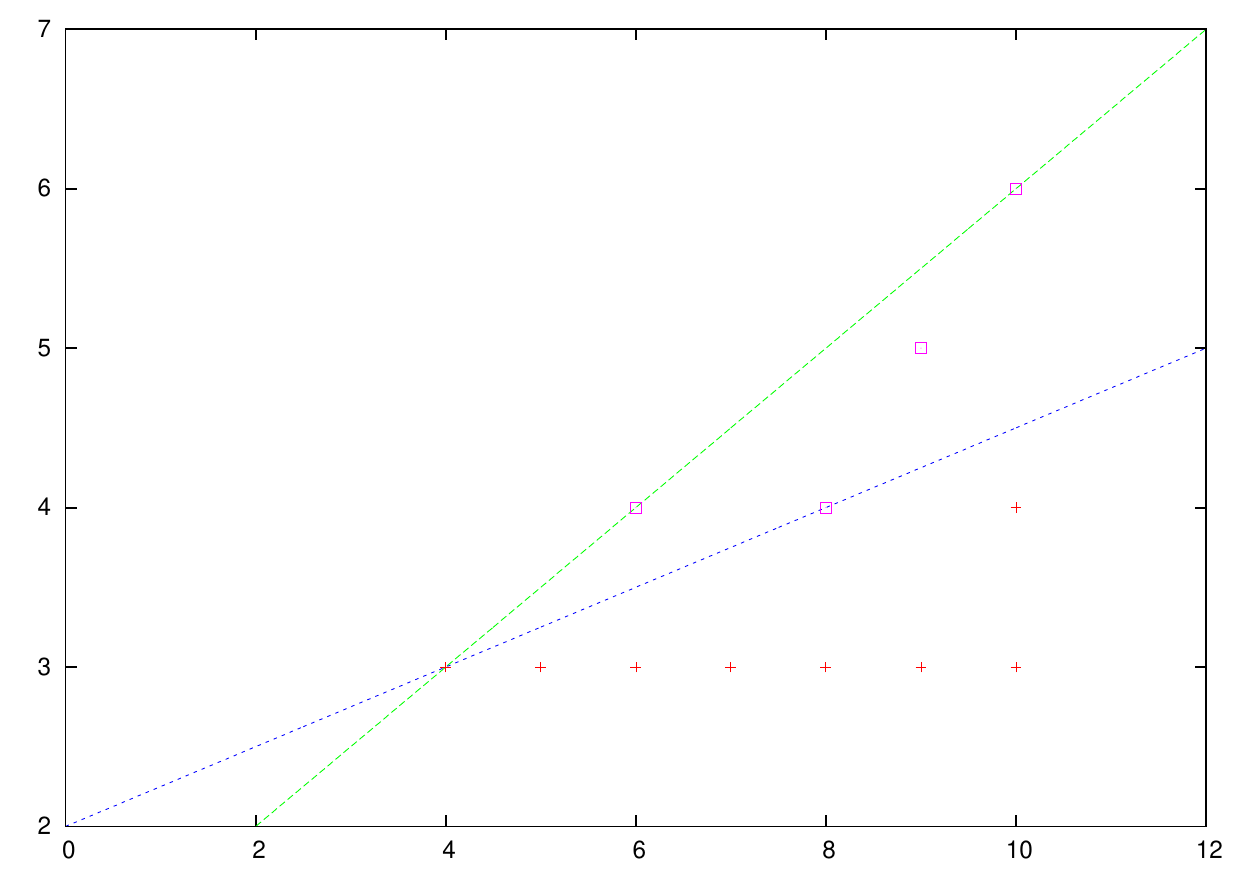}
&\includegraphics[width=0.47\hsize]{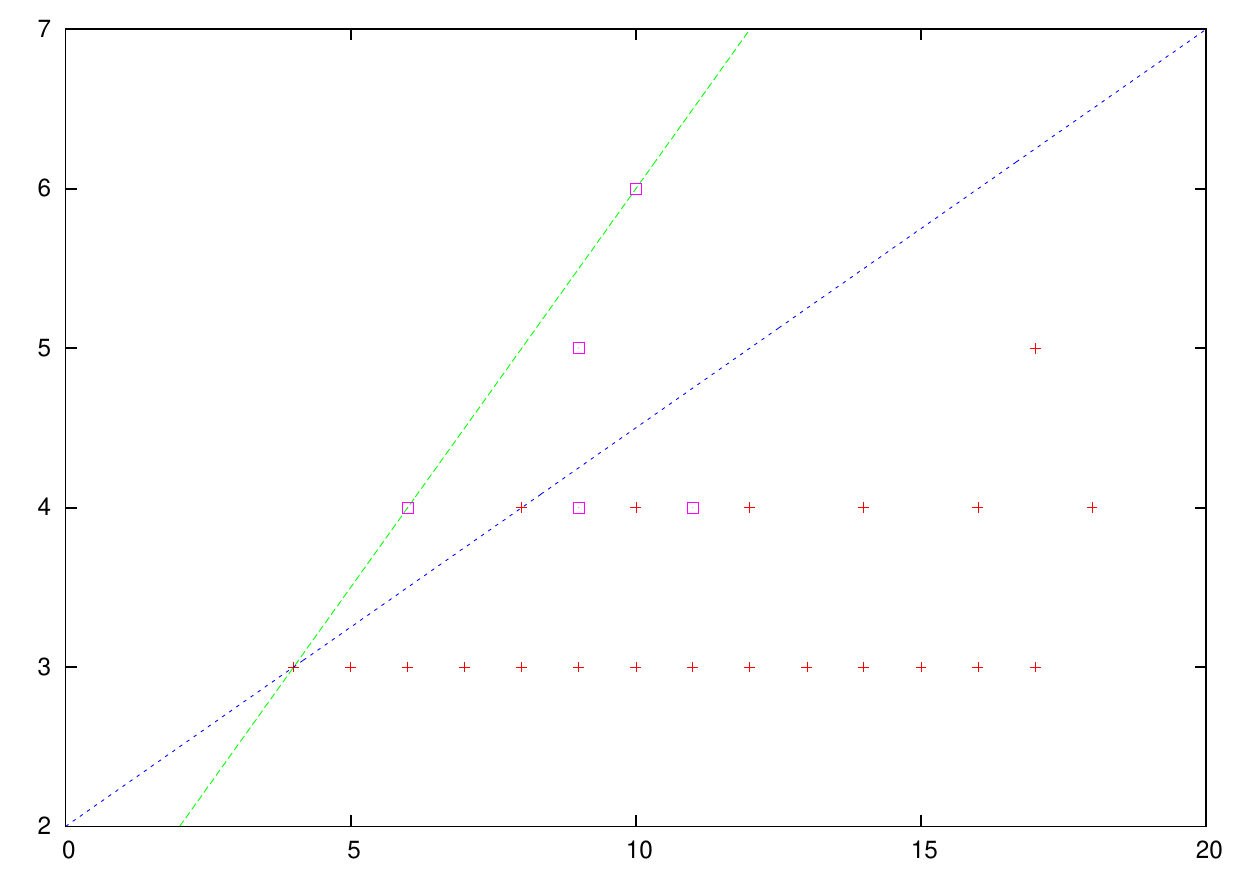}\\
$q=3$ & $q=4$\\[0.5ex]
\includegraphics[width=0.47\hsize]{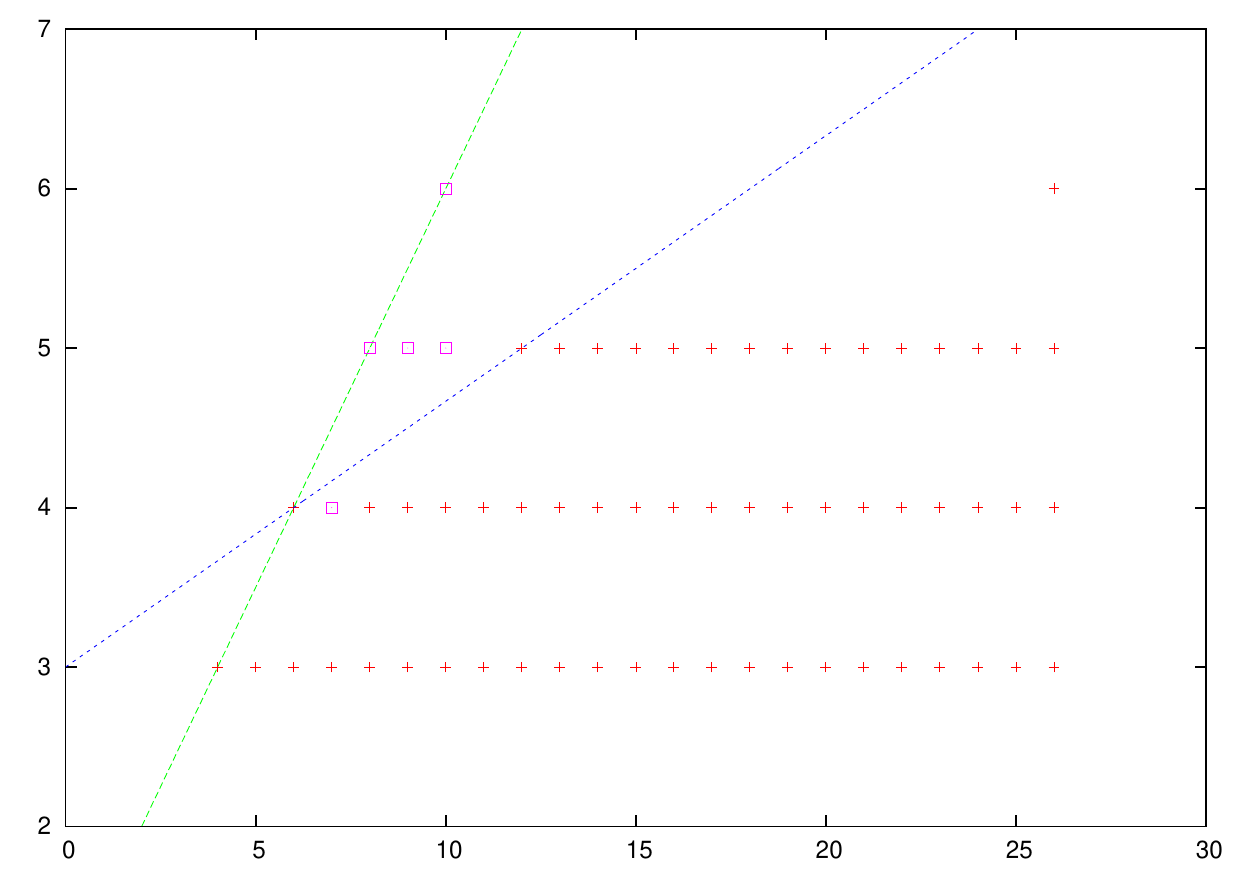}
&\includegraphics[width=0.47\hsize]{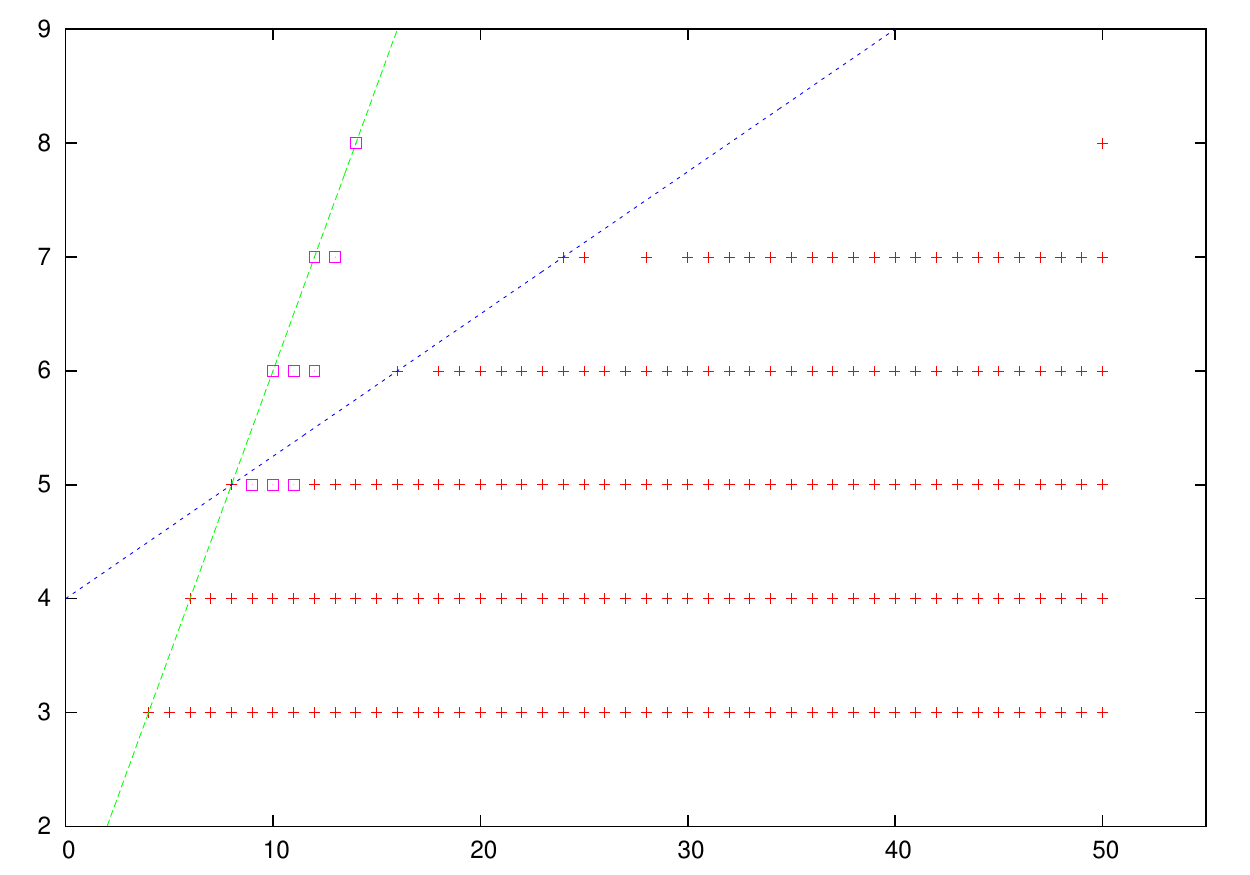}\\
$q=5$ & $q=7$
\end{tabular}
\caption{Minimum distance and length of QMDS codes for dimensions
  $3\le q\le 7$.}
\label{fig:results3}
\end{figure}

\begin{figure}[!ht]
\centering\scriptsize
\begin{tabular}{@{}cc@{}}
\includegraphics[width=0.47\hsize]{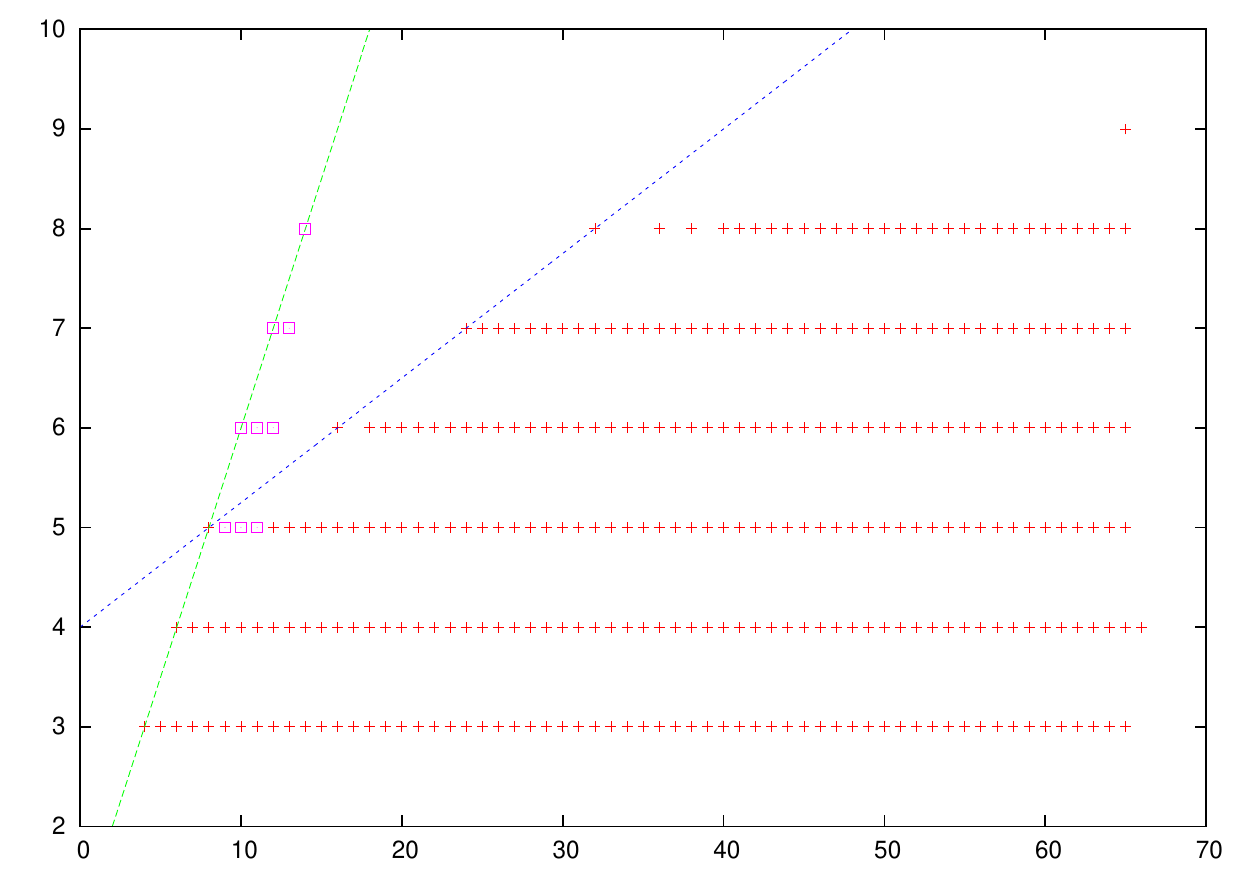}
&\includegraphics[width=0.47\hsize]{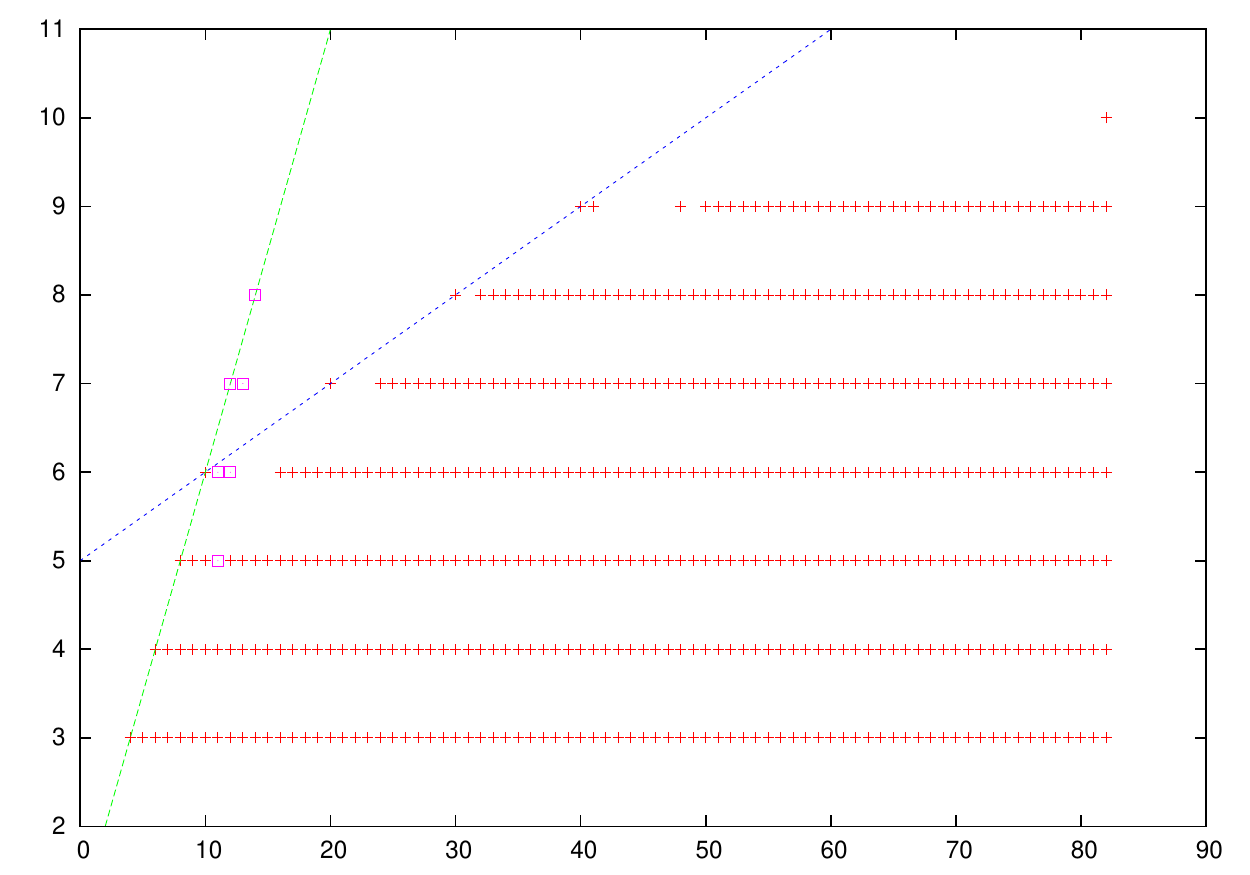}\\
$q=8$ & $q=9$\\[0.5ex]
\includegraphics[width=0.47\hsize]{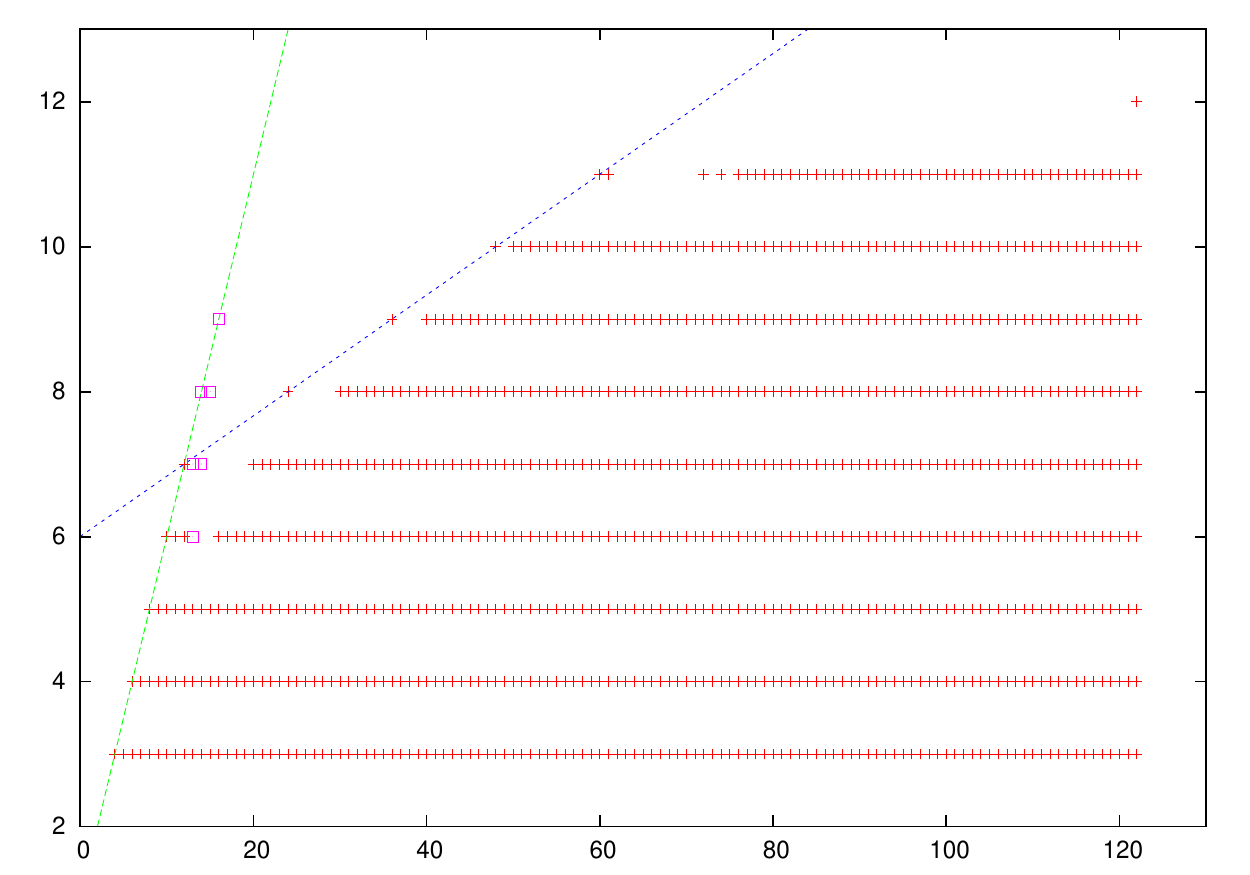}
&\includegraphics[width=0.47\hsize]{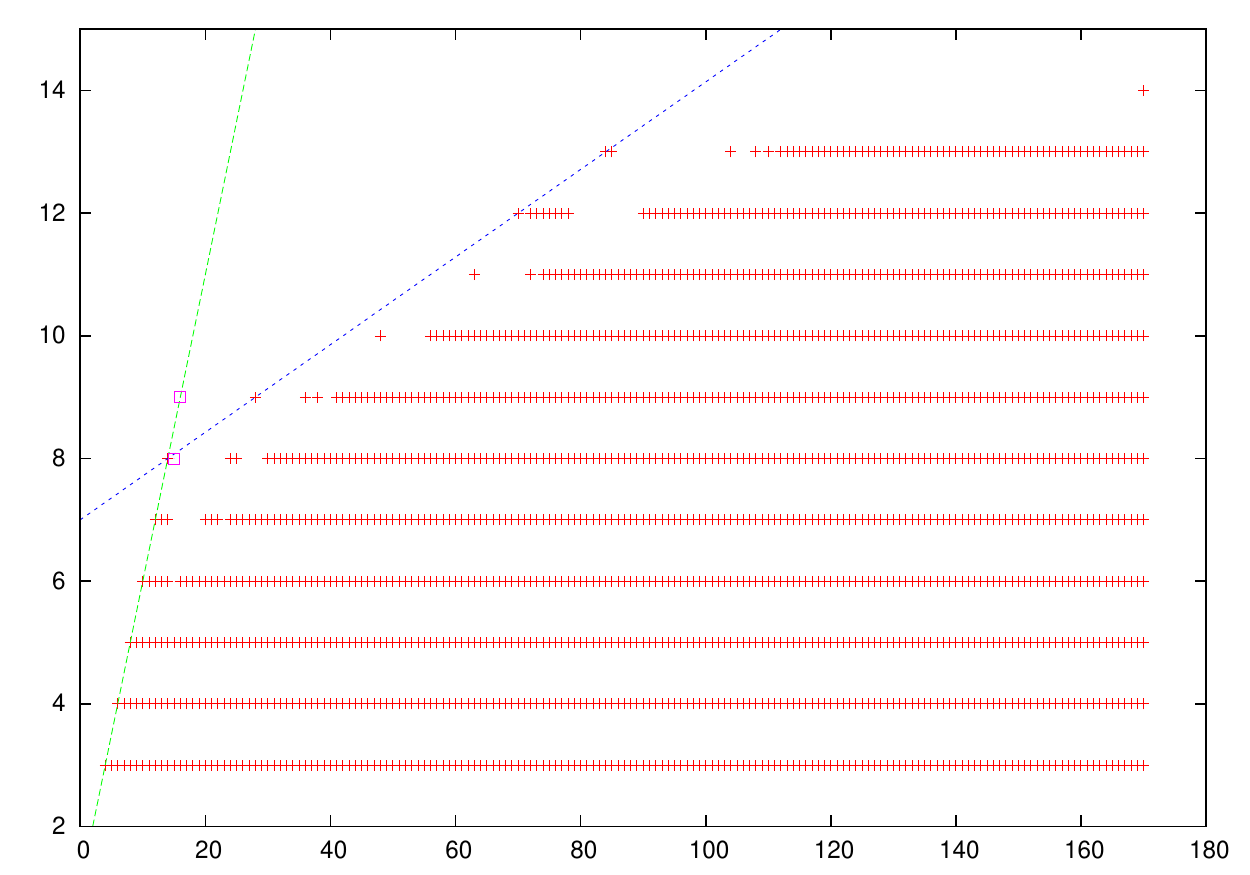}\\
$q=11$ & $q=13$\\[0.5ex]
\includegraphics[width=0.47\hsize]{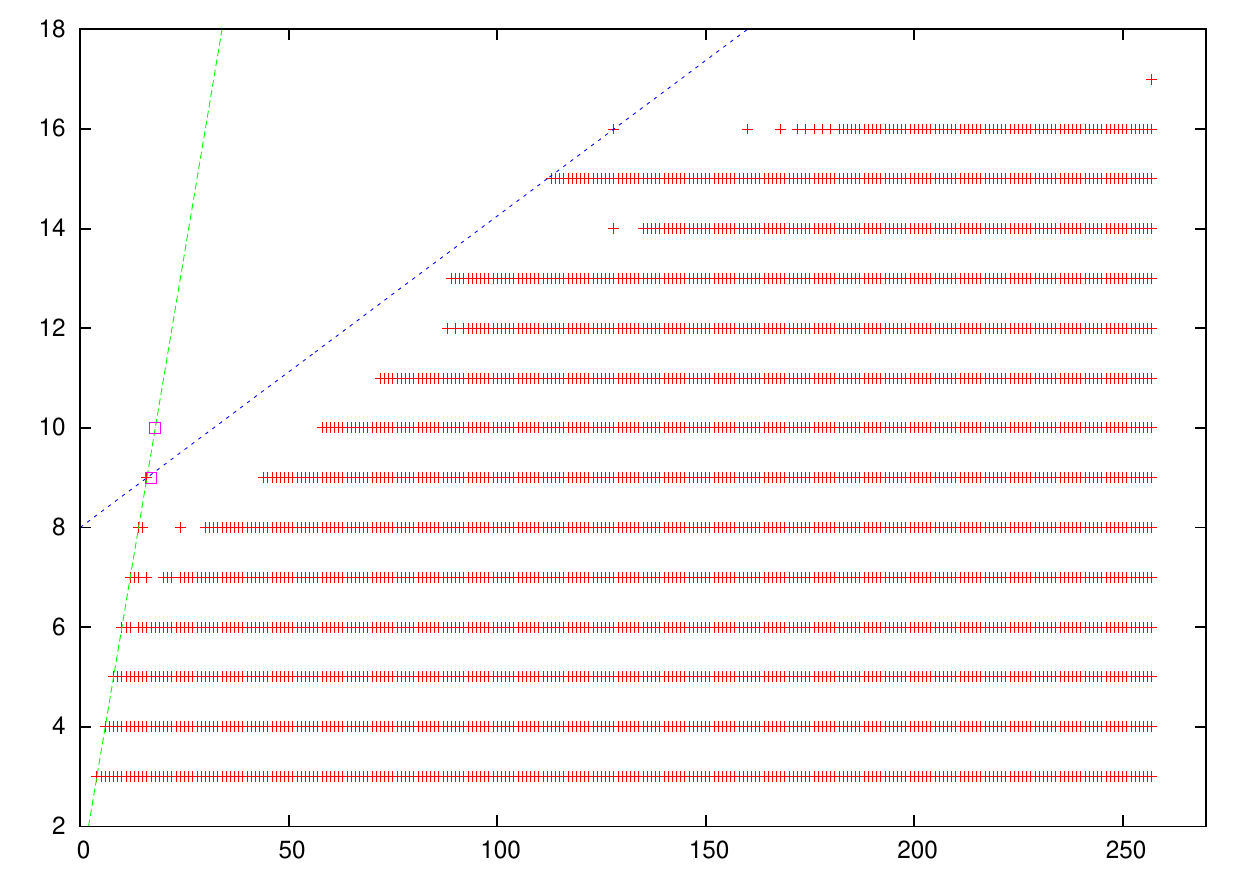}
&\includegraphics[width=0.47\hsize]{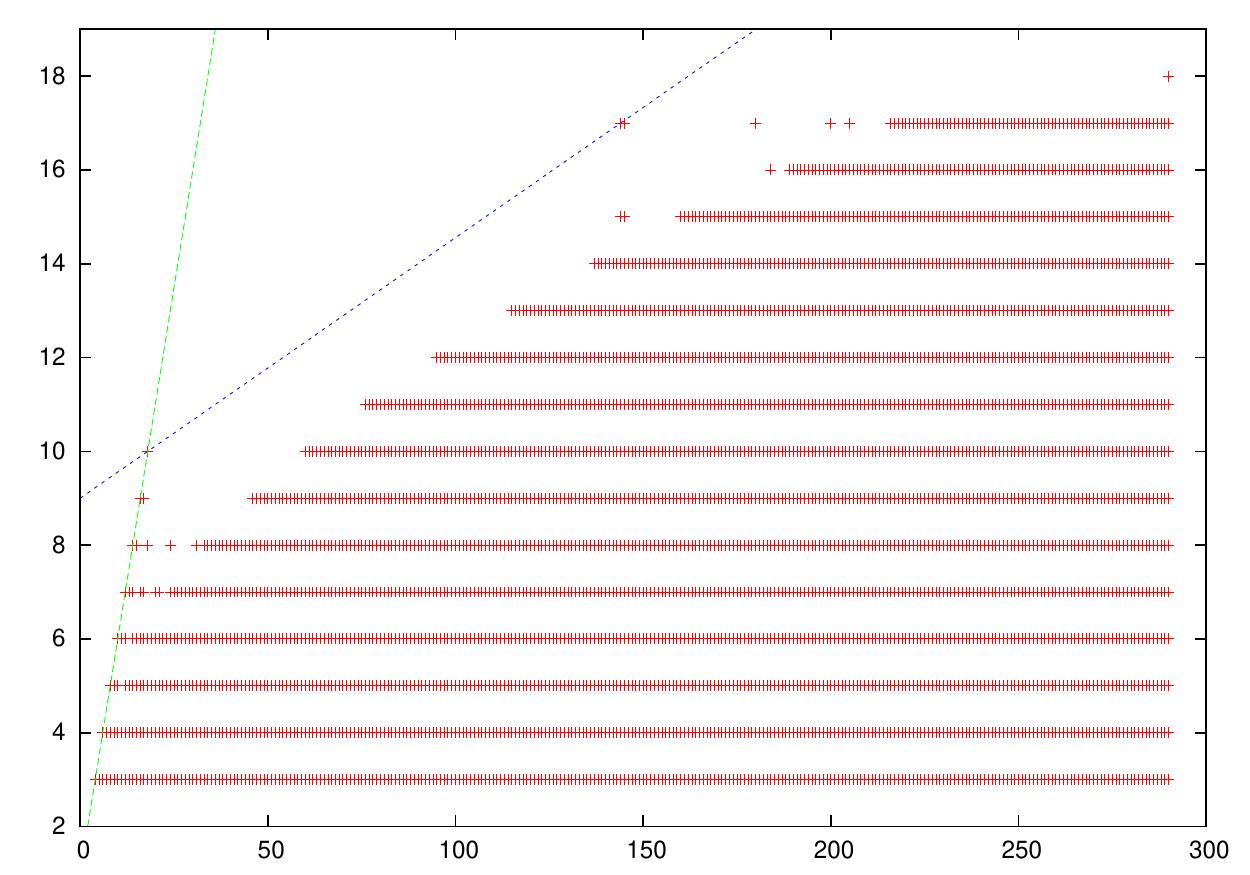}\\
$q=16$& $q=17$
\end{tabular}
\caption{Minimum distance and length of QMDS codes for dimensions
  $8\le q\le 17$.}
\label{fig:results}
\end{figure}

\begin{figure}[!ht]
\centering\scriptsize
\begin{tabular}{@{}cc@{}}
\includegraphics[width=0.47\hsize]{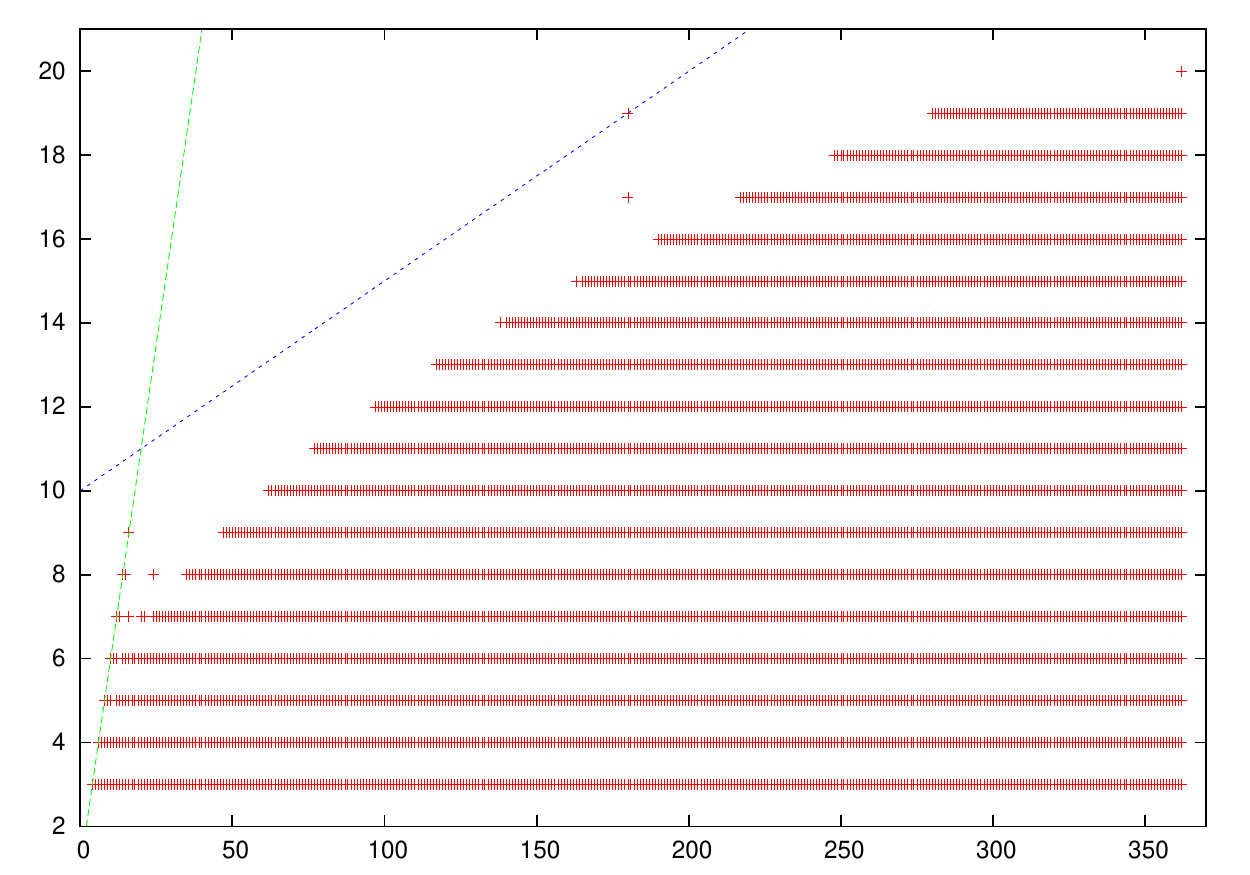}
&\includegraphics[width=0.47\hsize]{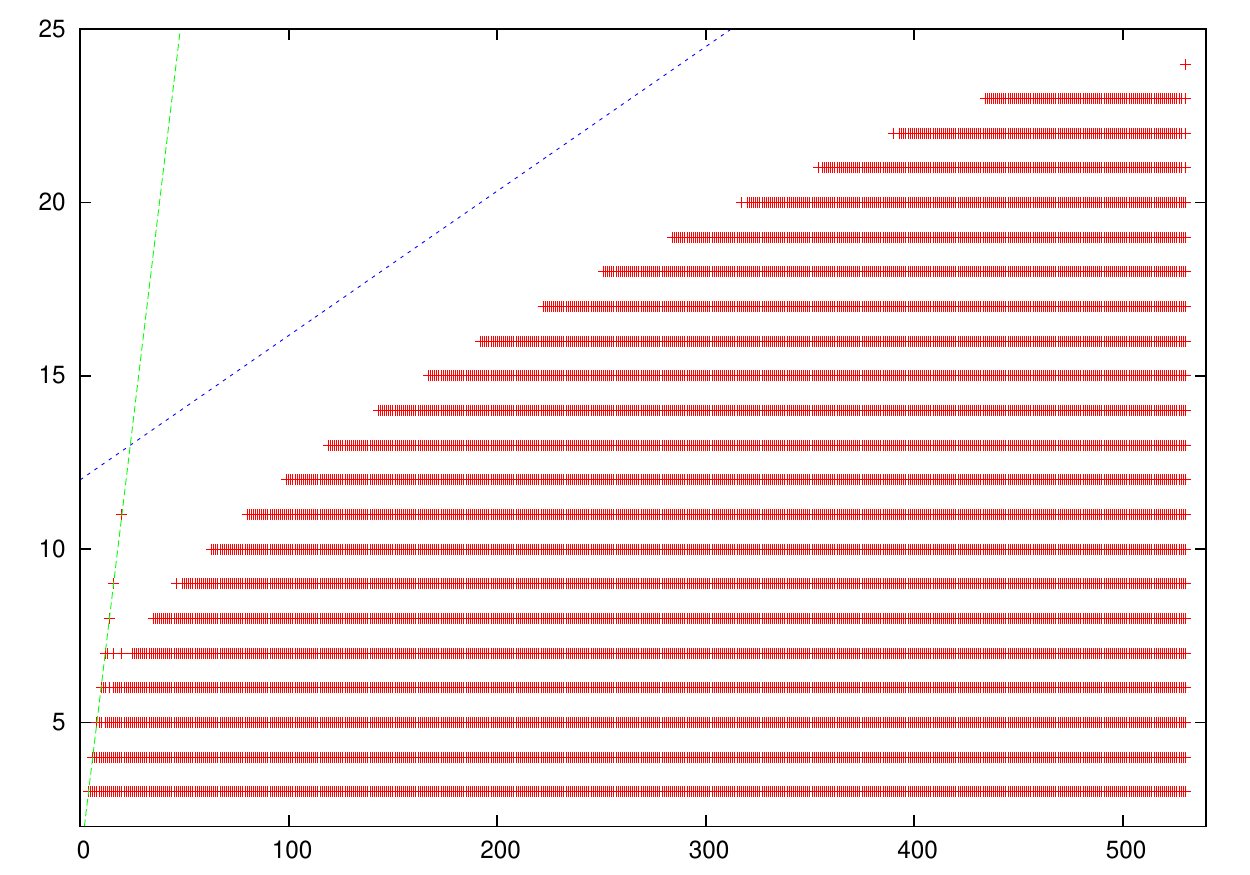}\\
$q=19$ & $q=23$\\[0.5ex]
\includegraphics[width=0.47\hsize]{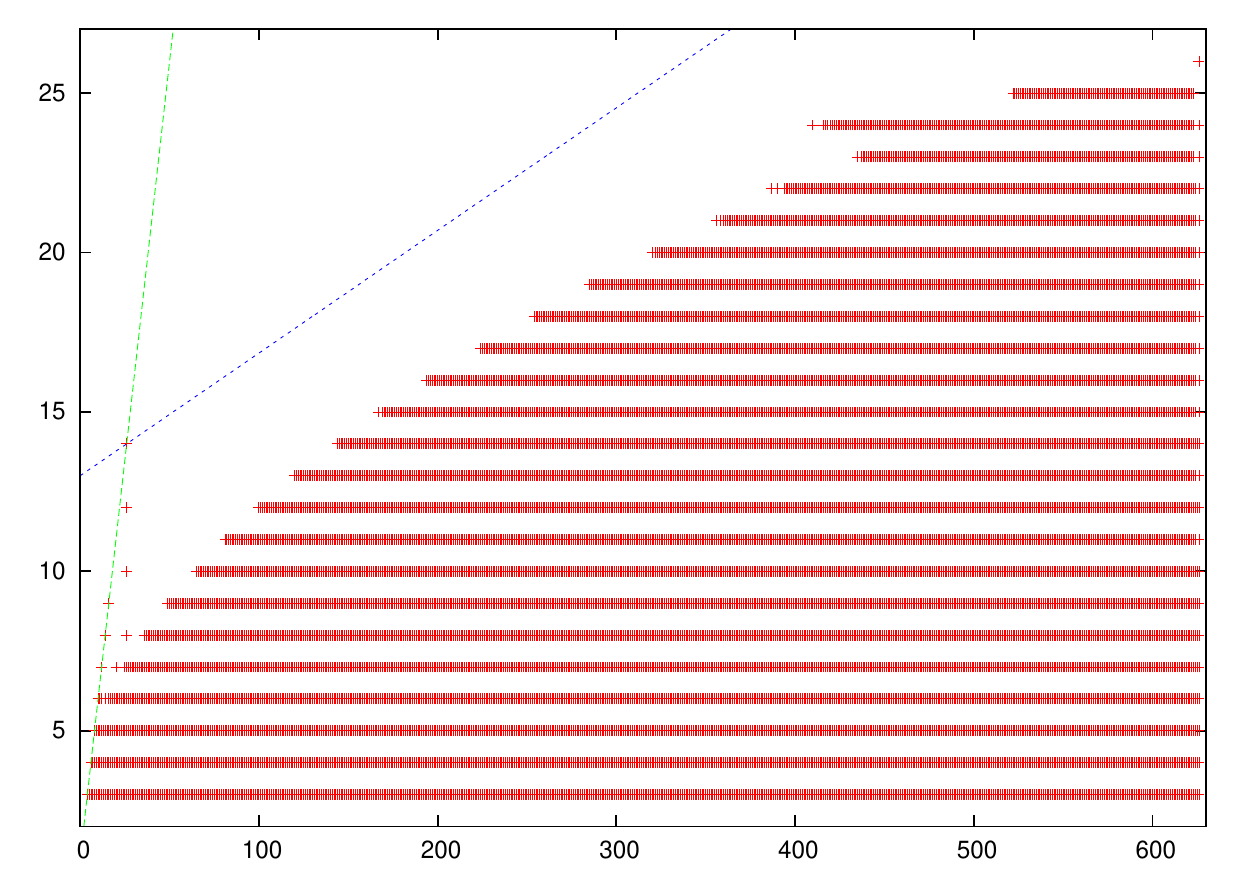}
&\includegraphics[width=0.47\hsize]{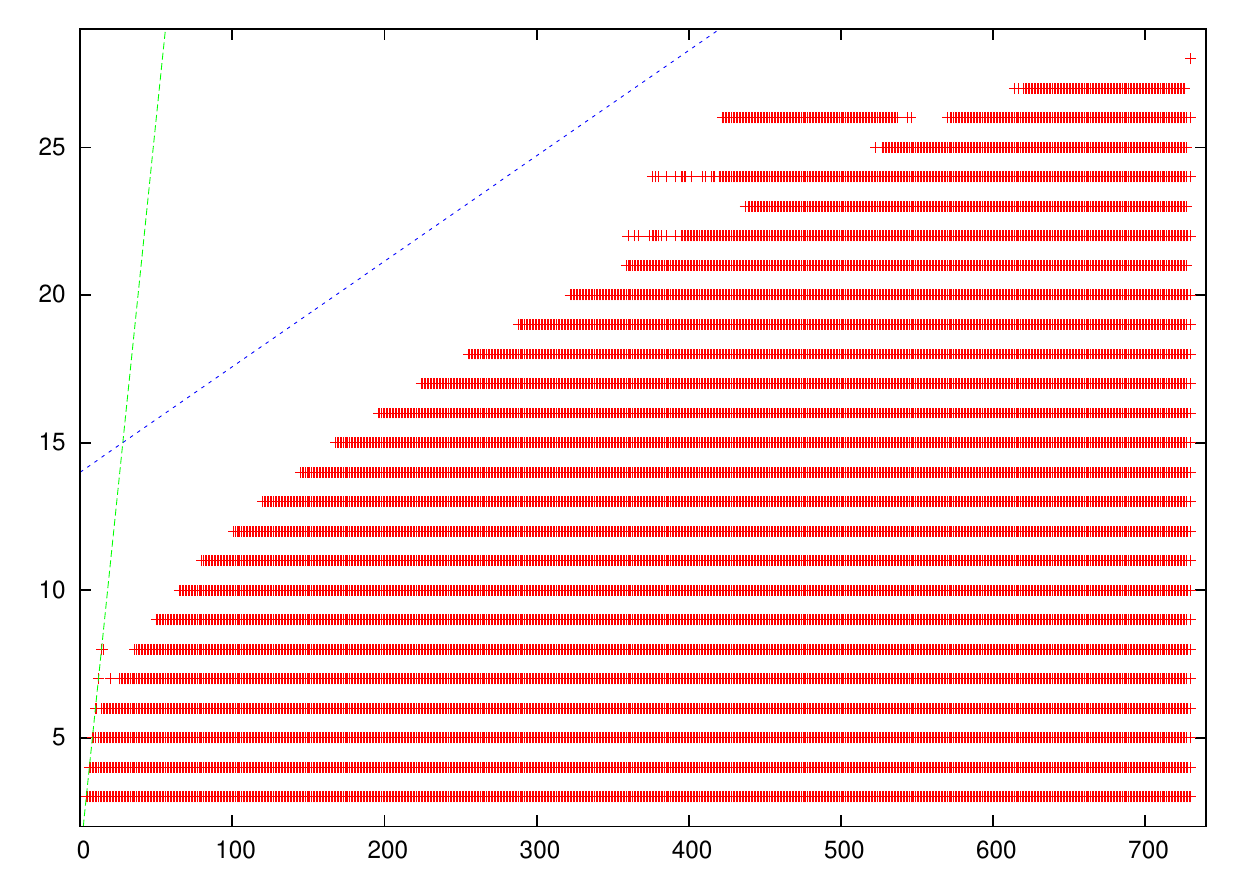}\\
$q=25$ & $q=27$\\[0.5ex]
\includegraphics[width=0.47\hsize]{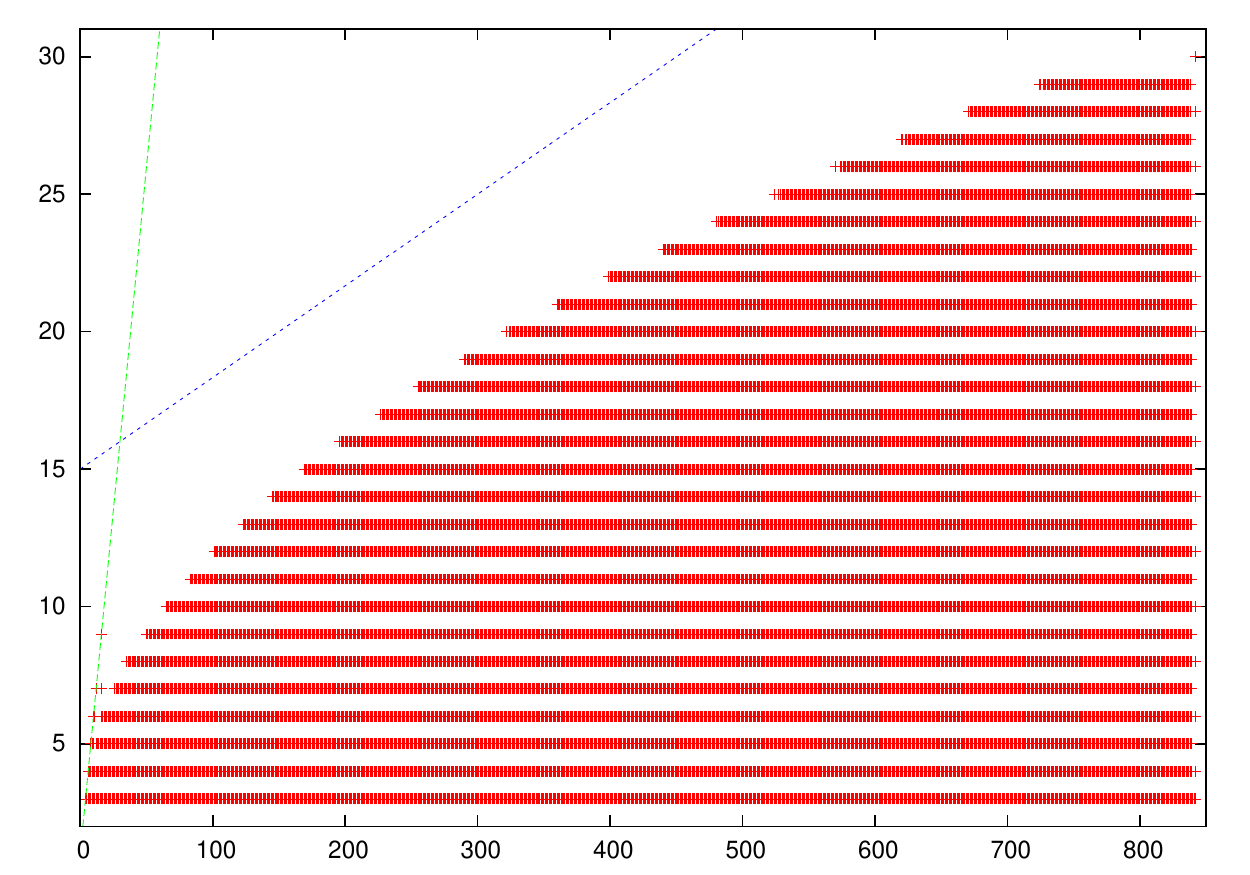}
&\includegraphics[width=0.47\hsize]{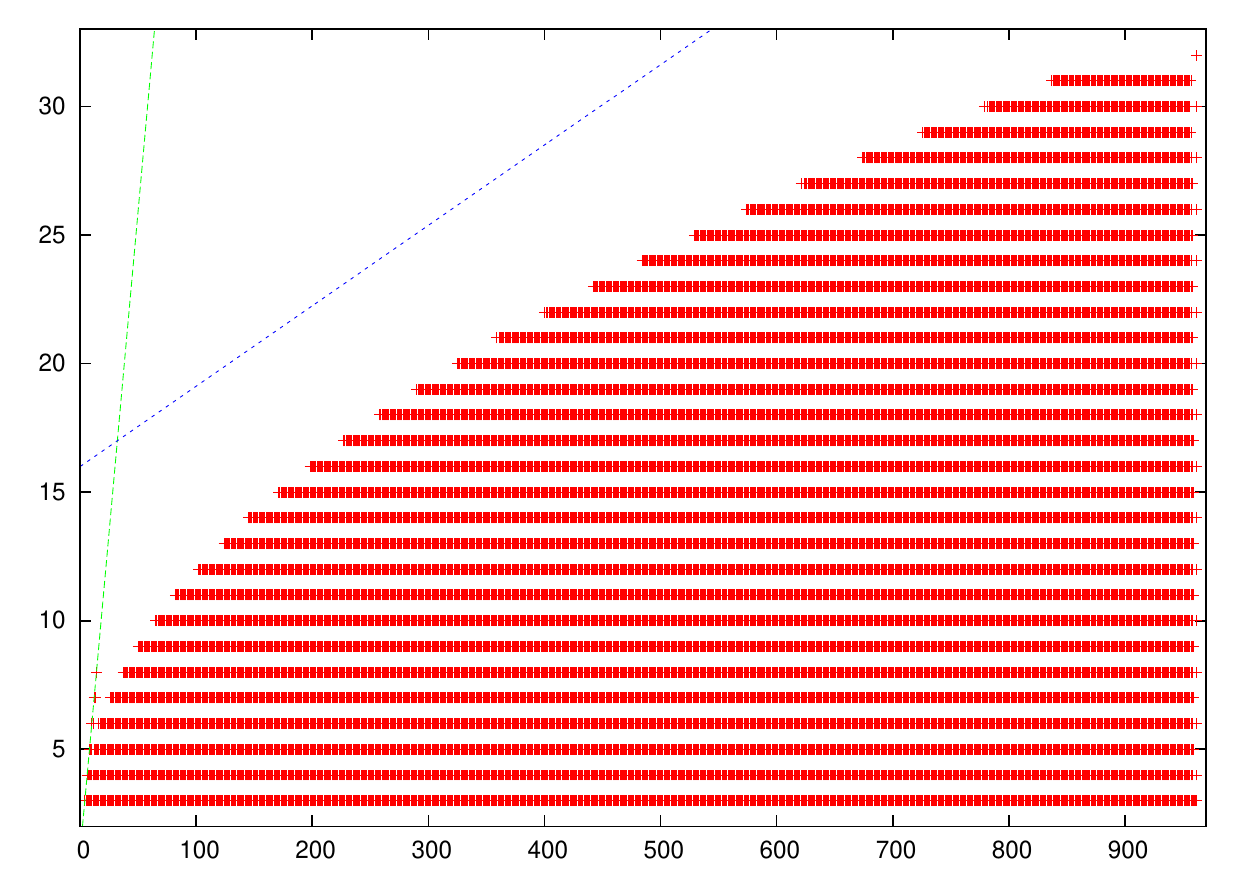}\\
$q=29$& $q=31$\\[0.5ex]
\includegraphics[width=0.47\hsize]{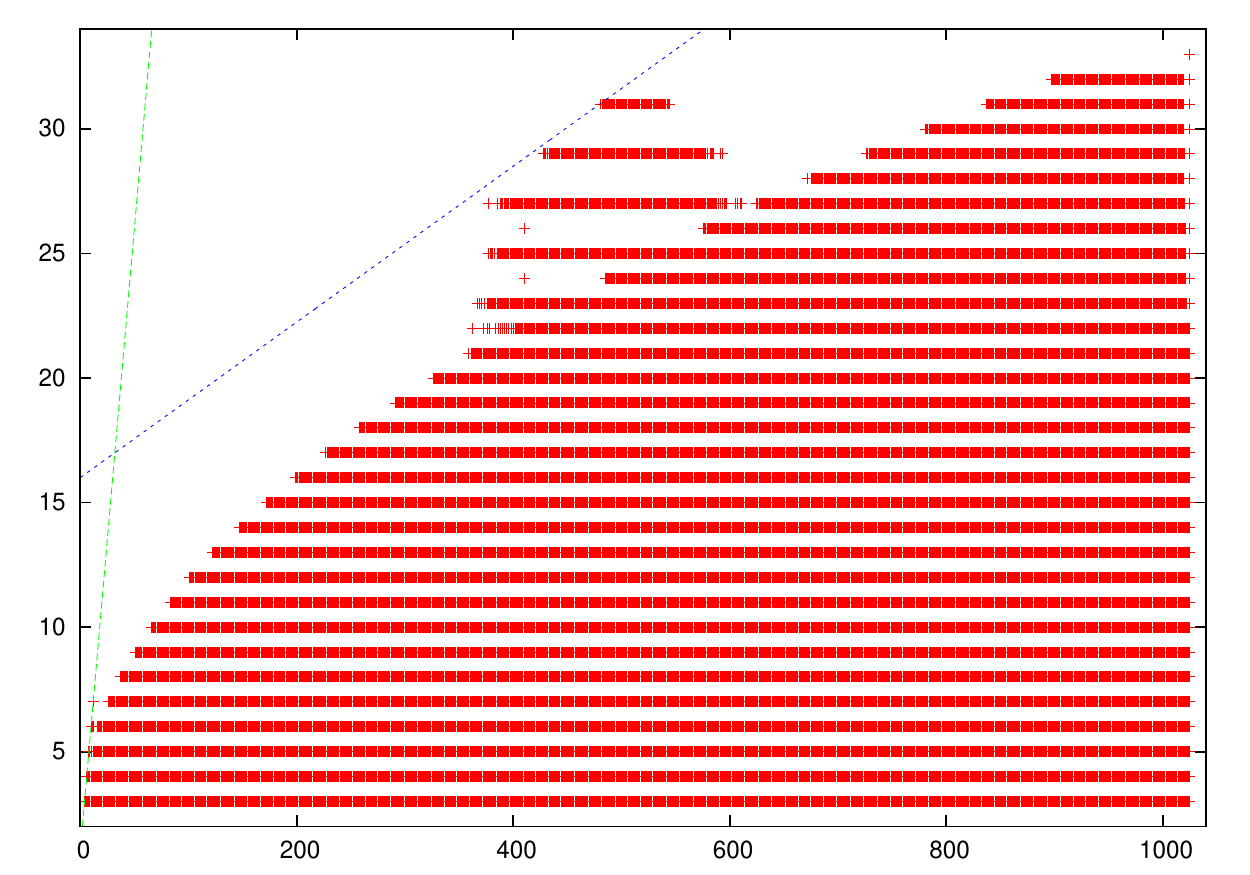}
&\includegraphics[width=0.47\hsize]{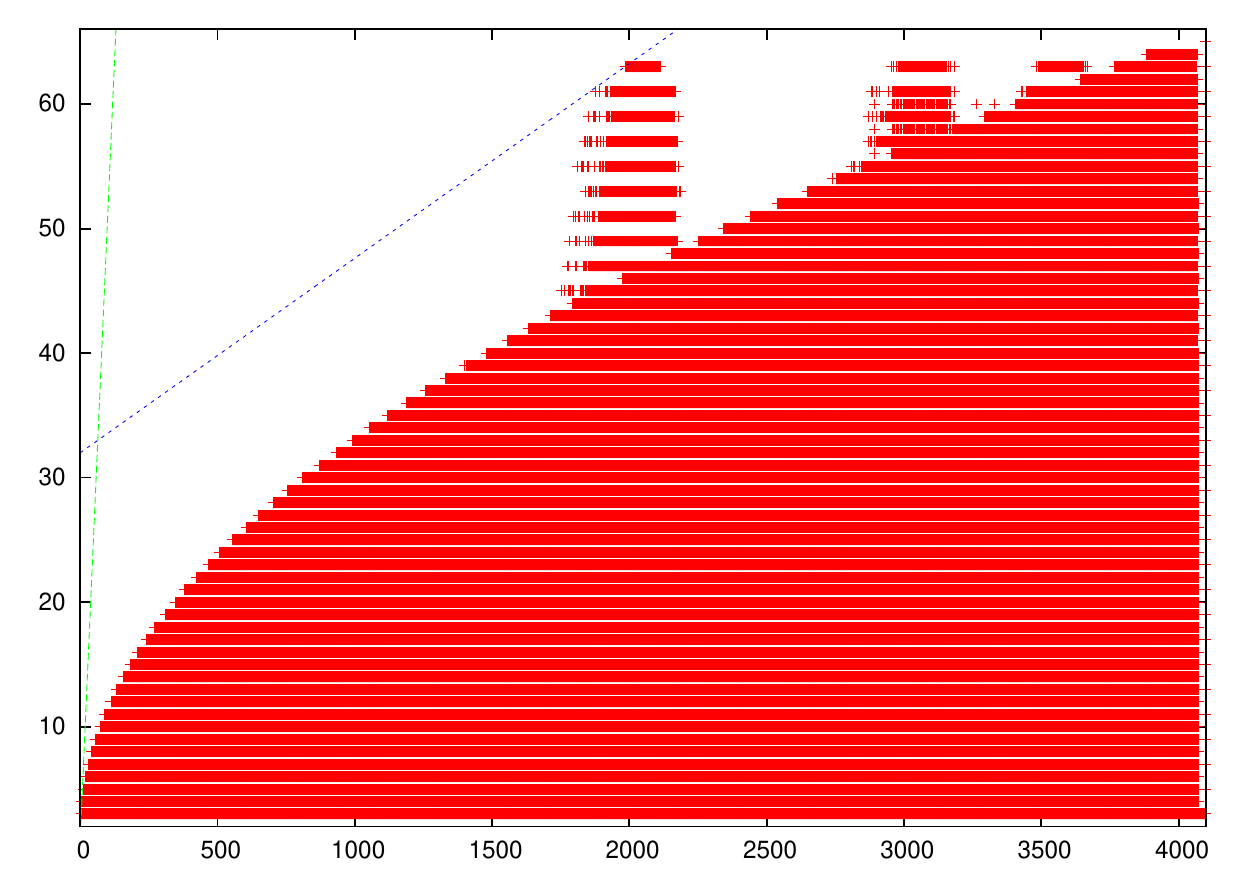}\\
$q=32$& $q=64$
\end{tabular}
\caption{Minimum distance and length of QMDS codes for dimensions
  $q\le 19 \le 32$ and $q=64$.}
\label{fig:results2}
\end{figure}

While we have complete information about the weight spectra of the
puncture codes for $q\le 8$, the information about the weights in
$P(C)$ for $q\ge 9$ is mainly based on sampling codewords.  In
Fig. \ref{fig:results} we plot the minimum distance ($y$-axis) against
the length of QMDS codes obtained by our construction for
$q=8,\ldots,17$.  We also plot the conjectured lower bounds on the
minimum weight of $P(C)$. The squares indicate QMDS codes that were
found by different methods.  Note that the steeper bound corresponds
to the quantum Singleton bound for $k=0$.  With increasing alphabet
size $q$, and hence increasing length $q^2+1$, it becomes more
difficult to find codewords of low weight in $P(C)$. For $q=16$, we
use subfield-subcodes over $\F_4$ and $\F_2$.  Fig. \ref{fig:results3}
and Fig. \ref{fig:results2} show the corresponding data for
$q=3,4,5,7$ and $q=19,\ldots,32,64$, respectively.  Again, for
$q=5^2,3^3,2^5,2^6$ we use subfield-subcodes of $P(C)$ as well.

\section{Conclusion}
In analogy to the conjecture for classical MDS codes, it has been
conjectured that the length of non-trivial quantum MDS codes is
bounded by $q^2+1$, or $q^2+2$ for very specific values \cite{KKKS06}.
Our results suggest that we can indeed find MQDS codes for all lengths
up to this bound.  Somewhat surprisingly, it seems that with
increasing minimum distance, it might become more difficult to find
QMDS codes below a certain length. Moreover, our construction is
restricted to QMDS codes of minimum distance $q+1$.  Nonetheless, we
found QMDS codes $[\![10,0,6]\!]_3$, $[\![9,1,5]\!]_3$, and
$[\![10,0,6]\!]_4$ with $d>q+1$.

We finally note that we can apply the technique of the puncture code
also to classical MDS codes of length $q^2-1$ or $q^2$ based on
(extended) Reed-Solomon codes. Obviously, this does not yield QMDS
codes of length $q^2+1$, but in terms of the weight spectra of the
codes $P(C)$, and hence the achievable lengths of QMDS codes, we get
essentially the same results.





%



\end{document}